\documentclass{ifacconf}

\usepackage{graphicx}      
\usepackage{natbib}        

\usepackage{graphicx}

\usepackage{amsmath}
\usepackage{amssymb}
\usepackage{amsfonts}
\usepackage{times}
\usepackage{mathptmx} 
\usepackage{datetime}

\newcommand{\sinc}{\mathrm{sinc}}
\newcommand{\tanc}{\mathrm{tanc}}
\newcommand{\tanhc}{\mathrm{tanhc}}

\DeclareMathAlphabet{\bit}{OML}{cmm}{b}{it}
\def\<{\leqslant}           
\def\>{\geqslant}           

\def\d{\partial}

\def\Re{\mathrm{Re}}   
\def\rprod{\mathop{\overrightarrow{\prod}}}

\def\cH{\mathcal{H}}   
\def\mR{{\mathbb R}}    
\def\mC{\mathbb{C}}    

\def\Tr{\mathrm{Tr}}       
\def\rT{{\rm T}}        
\def\bP{\mathbf{P}}    
\def\bE{\mathbf{E}}    

\def\re{{\rm e}}        
\def\rd{{\rm d}}        

\def\cL{\mathcal{L}}

\def\bD{{\mathbf D}}

\def\bJ{\mathbf{J}}

\def\x{\times}
\def\ox{\otimes}

\def\fF{\mathfrak{F}}

\def\fH{\mathfrak{H}}
\def\fS{\mathfrak{S}}
\def\fP{\mathfrak{P}}

\def\bH{{\mathbf H}}

\def\cF{\mathcal{F}}

\def\cK{\mathcal{K}}

\def\cG{\mathcal{G}}
\def\cI{\mathcal{I}}
\def\cP{\mathcal{P}}

\def\eps{\epsilon}

\def\Ups{\Upsilon}

\def\Res{\mathop{\mathrm{Res}}}    

\begin{document}
\begin{frontmatter}
\title{Frequency-Domain Computation of Quadratic-Exponential Cost Functionals
for Linear Quantum Stochastic  Systems\thanksref{footnoteinfo}}

\thanks[footnoteinfo]{This work is supported by the Air Force Office of Scientific Research (AFOSR) under agreement number FA2386-16-1-4065 and the Australian Research Council under grant DP180101805.}%
\author[IGV]{Igor G. Vladimirov$^*$,\qquad Ian R. Petersen$^*$,\qquad Matthew R. James}%

\address[IGV]{Research School of Electrical, Energy and Materials Engineering, College of Engineering and Computer Science,
Australian National University, Canberra, Acton, ACT 2601,
Australia (e-mail: igor.g.vladimirov@gmail.com, i.r.petersen@gmail.com, matthew.james@anu.edu.au).}

\begin{abstract}                
This paper is concerned with quadratic-exponential functionals (QEFs) as risk-sensitive performance criteria for linear quantum stochastic systems driven by multichannel bosonic fields. Such costs impose an exponential penalty on quadratic functions of the quantum system variables over a bounded time interval, and their minimization secures a number of robustness properties for the system. We use an integral operator representation  of the QEF, obtained recently, in order to compute its asymptotic infinite-horizon growth rate in the invariant Gaussian state when the stable system is driven by vacuum input fields.  The resulting frequency-domain formulas express the QEF growth rate in terms of two spectral functions associated with the real and imaginary parts of the quantum covariance kernel of the system variables. We also discuss the computation of the QEF growth rate using homotopy and contour integration techniques and provide two illustrations including a numerical example with a two-mode oscillator.
\end{abstract}


\end{frontmatter}

\section{Introduction}

Quantum-mechanical adaptation of quadratic-exponential cost functionals,  originating from classical risk-sensitive control \cite{BV_1985,J_1973,W_1981}, provides a relevant addition to the mean square optimality criteria for linear quantum stochastic systems.  Such systems, governed by linear quantum stochastic differential equations (QSDEs)  in the framework of the Hudson-Parthasarathy calculus \cite{HP_1984,P_1992,P_2015}, are the main subject of linear quantum systems theory \cite{NY_2017,P_2017} which is concerned with tractable models of open quantum dynamics. In particular, quadratic cost functionals and their minimization provide a natural way to quantify and improve the performance of observers in filtering problems in terms of the mean square discrepancy between the system variables and their estimates \cite{MJ_2012}.

The quadratic exponential functional (QEF) \cite{VPJ_2018b} (see also \cite{B_1996}), which, similarly to its classical predecessors,  is organised as the averaged exponential of an integral of a  quadratic form of the system variables over a bounded time interval, pertains to important higher-order properties of the quantum system.  One of them is related to the worst-case values of mean square costs \cite{VPJ_2018b} in the presence of quantum statistical uncertainty,  when the actual system-field state differs from its nominal model, but not ``too much'' in the sense of the quantum relative entropy \cite{OW_2010}. Another property is concerned with the tail distributions for the quantum system trajectories \cite{VPJ_2018a}, which corresponds to the classical  Cramer type large deviations bounds. These properties  involve the QEF in such a way that its minimization makes the behaviour of the open quantum system more robust and conservative. The resulting performance analysis and optimal control problems require methods for computing and  minimizing the QEF, which is different from its time-ordered exponential counterpart in the original quantum risk-sensitive control formulation  \cite{J_2004,J_2005}.

The development of methods for computing the QEF has been a subject of several recent publications which have developed Lie-algebraic techniques \cite{VPJ_2019a}, parametric randomization \cite{VPJ_2018c} and quantum Karhunen-Loeve  expansions \cite{VPJ_2019b,VJP_2019} for this purpose. These results have led to an integral operator representation of the QEF \cite{VPJ_2019c} for open quantum harmonic oscillators (OQHOs) in Gaussian quantum states \cite{P_2010}. In addition to its relevance to quantum risk-sensitive control,  the approach, which has been used in obtaining this representation, has  deep connections with operator exponential structures studied in mathematical physics and quantum probability (for example, in the context of operator algebras \cite{AB_2018}, moment-generating functions for quadratic Hamiltonians \cite{PS_2015} and the quantum L\'{e}vy area \cite{CH_2013,H_2018}).

The present paper employs the finite-horizon  representation of the QEF, mentioned above, and establishes an infinite-horizon  asymptotic growth rate of the QEF for invariant Gaussian states of stable OQHOs driven by vacuum input fields.  We represent the QEF growth rate in frequency domain through the Fourier transforms of the real and imaginary parts of the invariant quantum covariance kernel of the system variables. One of these matrix-valued spectral functions, coming from the two-point commutator kernel,  enters the frequency-domain formula in composition with trigonometric functions \cite{H_2008}. This affects the (otherwise meromorphic) structure of the function (whose logarithm is present in the integrand) in comparison with its classical counterpart in the $\cH_\infty$-entropy integral \cite{AK_1981,MG_1990}. We take into account this issue when considering a contour integration technique for evaluating the QEF growth rate and discuss the more complicated nature of singularities in the case of one-mode OQHOs with positive definite energy matrices. For general multimode OQHOs,  we obtain a differential equation for the QEF growth rate as a function of the risk sensitivity parameter, which leads to  a numerical algorithm for its computation, similar to the homotopy methods \cite{MB_1985}.

The paper is organised as follows.
Section~\ref{sec:sys} specifies the class of linear quantum stochastic systems under consideration.
Section~\ref{sec:QEF} describes the QEF as a finite-horizon  system performance criterion and revisits its integral operator representation in the Gaussian case.
Section~\ref{sec:freq}  obtains a frequency-domain formula for the infinite-horizon asymptotic growth rate of the QEF in terms of the system transfer function.
Section~\ref{sec:comp}  discusses the computation of the QEF growth rate using homotopy and contour integration techniques.
Section~\ref{sec:one} demonstrates the complicated nature of singularities of the integrand in the one-mode case.
Section~\ref{sec:num} provides a numerical example of computing the QEF growth rate
 for  a two-mode OQHO.
Section~\ref{sec:conc} makes concluding remarks and outlines further directions of research.

\section{Open quantum harmonic oscillators}
\label{sec:sys}

Let  $W_1(t), \ldots, W_m(t)$ be an even number of time-varying self-adjoint operators on a subspace $\fF_t$ of a symmetric Fock space $\fF$ \cite{P_1992}, which form a multichannel quantum Wiener process $W:=(W_k)_{1\< k \< m}$  and represent bosonic fields (we will often omit the time argument $t$ for brevity). The increasing family $(\fF_t)_{t\> 0}$   of these subspaces provides a filtration for the Fock space $\fF$  in accordance with its continuous tensor-product structure  \cite{PS_1972}. The quantum Wiener process $W$ satisfies the two-point canonical commutation relations (CCRs )
\begin{equation}
\label{WWcomm}
    [W(s), W(t)^\rT]
     := ([W_j(s), W_k(t)])_{1\< j,k\< m}
     =
    2i\min(s,t)J
\end{equation}
for all $s,t\>0$, where  $(\cdot)^\rT$ is the transpose (vectors are organised as columns unless indicated otherwise),    $[\alpha,\beta]:= \alpha\beta - \beta\alpha$ is the commutator of linear operators, and $i:= \sqrt{-1}$ is the imaginary unit. In (\ref{WWcomm}), use is also made of an orthogonal real  antisymmetric matrix
\begin{equation}
\label{JJ}
    J:=  \bJ \ox I_{m/2}
\end{equation}
(so that $J^2 = -I_m$),
where $\ox$ is the Kronecker product, $I_r$ is the identity matrix of order $r$, and
\begin{equation}
\label{bJ}
\bJ: = {\small\begin{bmatrix}
        0 & 1\\
        -1 & 0
    \end{bmatrix}}
\end{equation}
spans the one-dimensional subspace of antisymmetric matrices of order 2.  In addition to its relation to the second Pauli matrix  $-i\bJ$ \cite{S_1994}, this matrix also specifies the CCRs $[\vartheta, \vartheta^\rT] = i\bJ$ for the vector
\begin{equation}
\label{zeta}
    \vartheta:=
    {\small\begin{bmatrix}
        q\\
        p
    \end{bmatrix}}
\end{equation}
 of the quantum mechanical position and momentum operators $q$ and $p:= -i\d_q$ on the Schwartz space \cite{V_2002}. More complicated CCRs between quantum variables are obtained by using linear combinations of the conjugate position-momentum pairs as building blocks.

Such combinations are present in a multimode OQHO, which interacts with external bosonic fields (modelled by the   quantum Wiener process $W$) and is endowed with an even number of time-varying  self-adjoint quantum variables $X_1(t), \ldots, X_n(t)$ on the subspace $\fH_t:= \fH_0 \ox \fF_t$ of the system-field tensor-product space
\begin{equation}
\label{fH}
    \fH:= \fH_0 \ox \fF.
\end{equation}
Accordingly, $\fH_0$ is a complex separable Hilbert space for the action of the initial system variables $X_1(0), \ldots, X_n(0)$. At every moment of time,  the vector     $X:=(X_k)_{1\< k\< n}$ of system variables of the OQHO satisfies the CCRs
\begin{equation}
\label{XCCR}
    [X,X^\rT] = 2i \Theta
\end{equation}
as the Heisenberg infinitesimal form of the Weyl CCRs \cite{F_1989}, specified by a
constant real antisymmetric matrix $\Theta$ of order $n$, which is assumed to be nonsingular for what follows.   The evolution of the system variables is governed by a linear QSDE
\begin{equation}
\label{dX}
    \rd X = AX \rd t + B\rd W,
\end{equation}
driven by the quantum Wiener process $W$. In accordance with the structure of the system-field interaction model in the quantum stochastic calculus \cite{HP_1984,P_1992,P_2015}, the matrices $A \in \mR^{n\x n}$, $B\in \mR^{n\x m}$ are parameterised as
\begin{equation}
\label{AB}
    A = 2\Theta (R + M^\rT JM),
     \qquad
     B = 2\Theta M^\rT
\end{equation}
by the energy and coupling matrices $R = R^\rT \in \mR^{n\x n}$, $M \in \mR^{m\x n}$ which specify the system Hamiltonian
\begin{equation}
\label{H}
    H:= \tfrac{1}{2} X^\rT R X
\end{equation}
and the vector $MX$ of $m$ system-field coupling operators, with the matrix $J$ given by (\ref{JJ}).  Due to the parameterisation (\ref{AB}), the matrices $A$, $B$ satisfy the physical realizability (PR) condition \cite{JNP_2008}
\begin{equation}
\label{PR}
    A \Theta + \Theta A^\rT +  BJB^\rT = 0,
\end{equation}
which is equivalent to the conservation of the CCR matrix $\Theta$ in (\ref{XCCR}) in time. For what follows, the OQHO is assumed to be stable in the sense of $A$ being Hurwitz. In this case,  $\Theta$ is a unique solution of (\ref{PR}) as an algebraic Lyapunov equation (ALE) and is given by $\Theta = \int_0^{+\infty} \re^{tA} BJB^\rT\re^{tA^\rT} \rd t$.

\section{Quadratic-exponential cost functional}
\label{sec:QEF}

Feedback connections of  linear quantum stochastic systems, arising in quantum control and filtering settings \cite{NJP_2009,MJ_2012,ZJ_2012}, are also  organised as OQHOs, described in Section~\ref{sec:sys}.   For a  given but otherwise arbitrary time horizon $T>0$,  the performance of such a system over the time interval $[0,T]$  can be described in the risk-sensitive  framework in terms of the QEF \cite{VPJ_2018a}
\begin{equation}
\label{Xi}
    \Xi_{\theta,T}
    :=
    \bE \re^{\frac{\theta}{2} Q_T }
\end{equation}
as a cost functional to be minimised. Here,
$\bE \zeta := \Tr(\rho \zeta)$ is the quantum expectation over an underlying density operator $\rho$ on the system-field space $\fH$ in (\ref{fH}). The risk sensitivity parameter $\theta>0$ in (\ref{Xi}), divided by 2 for convenience,  specifies the severity of exponential penalty imposed on the positive semi-definite self-adjoint quantum variable
\begin{equation}
\label{Q}
    Q_T
    :=
    \int_0^T
    X(t)^\rT \Pi X(t)\rd t
    =
    \int_0^T
    Z(t)^\rT Z(t)\rd t,
\end{equation}
which
 depends quadratically on the system variables in (\ref{dX}) over the time interval $[0,T]$. This dependence  is parameterised by a real positive definite symmetric matrix $\Pi$ of order $n$ which relates an auxiliary quantum process $Z$ to the system variables by
 \begin{equation}
 \label{ZX}
   Z:= S X,
   \qquad
   S:= \sqrt{\Pi}.
 \end{equation}
 In fact, $Z$ consists of $n$ system variables of an OQHO with appropriately
 transformed matrices $S\Theta S$, $S^{-1}RS^{-1}$, $M S^{-1}$, $SAS^{-1}$, $SB$ in (\ref{XCCR})--(\ref{AB}) in view of the symmetry $S=S^\rT$.
This transformation preserves the nonsingularity of the CCR matrix $\Theta$ and the Hurwitz property of the dynamics matrix $A$. The process $Z$ satisfies the two-point CCRs \cite{VPJ_2018a}
\begin{equation}
\label{ZZcomm}
    [Z(s), Z(t)^\rT]
    =
    2i\Lambda(s-t),
    \qquad
    s,t\>0,
\end{equation}
with
\begin{equation}
\label{Lambda}
    \Lambda(\tau)
     :=
    \left\{
    {\small\begin{matrix}
    S \re^{\tau A}\Theta S& {\rm if}\  \tau\> 0\\
    S \Theta\re^{-\tau A^{\rT}}S & {\rm if}\  \tau< 0\\
    \end{matrix}}
    \right.
    =
    -\Lambda(-\tau)^\rT,
\end{equation}
from which the one-point CCR matrix of $Z$  is recovered as $S\Theta S = \Lambda(0)$. The two-point CCR kernel (\ref{Lambda}) gives rise to a skew self-adjoint integral operator $\cL_T : f\mapsto g$  which acts on the Hilbert space $L^2([0,T],\mC^n)$ of square integrable $\mC^n$-valued functions on $[0,T]$ as
\begin{equation}
\label{cL}
    g(s)
    :=
    \int_0^T
    \Lambda(s-t) f(t)
    \rd t,
    \qquad
    0\< s \< T.
\end{equation}
Note that the commutation structure of the process $Z$ in (\ref{ZZcomm}), (\ref{Lambda}) (and the related operator (\ref{cL})) do not depend on a particular system-field state $\rho$.

In what follows, we will be concerned with the case when the stable OQHO under consideration is driven by vacuum fields. By an appropriate modification of the results of \cite{VPJ_2018a}, the system variables have a unique  invariant multipoint zero-mean Gaussian quantum state in this case. This property is inherited by the process $Z$ in (\ref{ZX}). The corresponding  two-point quantum covariance function
\begin{equation}
\label{EZZ}
\bE(Z(s)Z(t)^\rT)
  =
  P(s-t) + i\Lambda(s-t),
  \qquad
  s,t\> 0
\end{equation}
(with the imaginary part (\ref{Lambda}) irrespective of the quantum state) has the real part
\begin{equation}
\label{P}
    P(\tau)
    =
    \left\{
    {\small\begin{matrix}
    S\re^{\tau A}\Sigma S& {\rm if}\  \tau\> 0\\
    S \Sigma \re^{-\tau A^\rT}S & {\rm if}\  \tau < 0
    \end{matrix}}
    \right.
    =
    P(-\tau)^\rT,
    \qquad
    \tau \in \mR.
\end{equation}
Here, the real positive semi-definite symmetric matrix $\Sigma:= \Re \bE (XX^\rT)$ of order $n$ describes the invariant one-point statistical correlations of the system variables and satisfies the ALE
  $A \Sigma + \Sigma A^\rT + BB^\rT=0$.
  The kernel (\ref{P}) specifies a positive semi-definite self-adjoint integral  operator $\cP_T : f\mapsto g$ acting on $L^2([0,T], \mC^n)$ as
\begin{equation}
\label{cP}
    g(s):= \int_0^T P(s-t)f(t)\rd t,
    \qquad
    0\< s \< T.
\end{equation}
Moreover, the self-adjoint operator $\cP_T +i\cL_T $ on $L^2([0,T], \mC^n)$ is positive semi-definite, which is a stronger property than $\cP_T  \succcurlyeq 0$. Also note that both $\cP_T $ and $\cL_T $ are compact operators \cite{RS_1980}. Application of appropriately  modified results of \cite{VPJ_2019c} to the OQHO in the invariant multipoint Gaussian quantum state allows the QEF (\ref{Xi}) to be  computed as
\begin{equation}
\label{lnXi}
  \ln \Xi_{\theta,T}
    =
    -  \tfrac{1}{2}
  \Tr (\ln\cos (\theta\cL_T ) + \ln (\cI - \theta \cP_T \cK_{\theta,T}   )).
\end{equation}
Here,
\begin{equation}
\label{cK}
    \cK_{\theta,T}
    :=
    \tanhc(i\theta \cL_T ) = \tanc (\theta\cL_T )
\end{equation}
is a positive definite self-adjoint operator on $L^2([0,T],\mC^n)$, where $\tanhc z := \tanc (-iz)$ is a hyperbolic version of the function $\tanc z := \frac{\tan z}{z}$ extended  by continuity to $1$ at $z=0$. Note that $\cK_{\theta,T}  $ is a nonexpanding operator in the sense that $\cK_{\theta,T}   \preccurlyeq \cI$, with $\cI$ the identity operator on  $L^2([0,T],\mC^n)$. With $\cP_T \cK_{\theta,T}  $ being a compact operator (isospectral to the positive semi-definite self-adjoint operator $\sqrt{\cK_{\theta,T}  } \cP_T  \sqrt{\cK_{\theta,T}  }$), the representation (\ref{lnXi}) is valid under the condition
\begin{equation}
\label{spec}
    \theta \lambda_{\max}(\cP_T \cK_{\theta,T}  ) < 1,
\end{equation}
where $\lambda_{\max}(\cdot)$ is the largest eigenvalue. The representation (\ref{lnXi}) is obtained by applying the results of \cite{VPJ_2019c} to the process $Z$ in (\ref{Q}), (\ref{ZX})  using its quantum Karhunen-Loeve expansion over an orthonormal eigenbasis of the operator $\cL_T $ in (\ref{cL}), provided it has no zero eigenvalues. The latter property is inherited by $Z$ from the system variables  under the sufficient condition \cite[Theorem~1]{VPJ_2019c}
\begin{equation}
\label{BJB}
  \det (BJB^\rT) \ne 0,
\end{equation}
with $J$, $B$ given by (\ref{JJ}), (\ref{AB}).
Indeed, the corresponding condition $\det (SBJB^\rT S)\ne 0$ for the process $Z$ is equivalent to (\ref{BJB}) since the matrix $S$ in (\ref{ZX}) is nonsingular.

\section{QEF growth rate in the frequency domain}
\label{sec:freq}

The representation (\ref{lnXi}) employs ``trace-analytic'' \cite{VP_2010} functionals of operators in the sense that
\begin{equation}
\label{lnXi1}
  \ln \Xi_{\theta,T}
    =
    -  \tfrac{1}{2}
  \Tr (\varphi(\theta \cP_T \cK_{\theta,T}   ) + \psi(\theta\cL_T )),
\end{equation}
where
\begin{equation}
\label{phipsi}
  \varphi(z):= \ln(1-z),
  \qquad
  \psi(z):= \ln \cos z,
  \qquad
  z \in \mC,
\end{equation}
are holomorphic functions whose domains contain the spectra of the operators $\theta \cP_T \cK_{\theta,T}   $ (under the condition (\ref{spec})) and $\theta\cL_T $, at which these functions are evaluated.

We will now take into account the dependence of the   operators $\cP_T $, $\cL_T $ in (\ref{cP}), (\ref{cL}) (and the related operator $\cK_{\theta,T}  $ in (\ref{cK})) on the time horizon  $T>0$. Each of them is organised as an integral operator $\cF_T$ on $L^2([0,T],\mC^n)$ whose kernel $F_T: [0,T]^2\to \mC^{n\x n}$ is obtained from a function $f: \mR\to \mC^{n \x n}$ (a shift-invariant kernel) as $F_T(s,t):= f(s-t)$ for all  $0\< s,t\< T$. Accordingly, the composition $\cH_T:= \cF_T \cG_T$ of such integral operators with kernel functions $F_T, G_T: [0,T]^2\to \mC^{n\x n}$, generated by $f, g: \mR \to \mC^{n\x n}$,  is an integral operator whose kernel is an appropriately constrained convolution  $H_T(s,t):= \int_0^T F_T(s,u)G_T(u,t)\rd u = \int_0^T f(s-u)g(u-t)\rd u$ of the functions $f$, $g$ for all $0\< s,t\< T$. If the operator $\cH_T$ is of trace class \cite{RS_1980}, then  \cite{B_1988}
$$
    \Tr \cH_T
    =
    \int_0^T \Tr H_T(t,t)\rd t
    =
    \int_{[0,T]^2}
    \Tr (f(s-t)g(t-s))
    \rd s \rd t.
$$
This relation extends to the rightward-ordered product $\cF_T := \rprod_{k=1}^r \cF_T^{(k)}$ of any number $r$ of such operators  with the kernel functions $F_T^{(k)}: [0,T]^2 \to \mC^{n\x n}$, generated by $f_k: \mR \to \mC^{n\x n}$ as above, with $k=1, \ldots, r$, so that if the operator $\cF_T$ is of trace class, then
\begin{equation}
\label{TrFT}
    \Tr \cF_T
    =
    \int_{[0,T]^r}
    \Tr
    \rprod_{k=1}^r
    f_k (t_k - t_{k+1})
    \rd t_1\x \ldots \x \rd t_r,
\end{equation}
where $t_{r+1}:= t_1$.
Application of \cite[Lemma~6, Appendix~C]{VPJ_2018a} to (\ref{TrFT}) leads to
\begin{equation}
\label{lim}
    \lim_{T\to +\infty}
    \big(
        \tfrac{1}{T}
        \Tr \cF_T
    \big)
    =
    \tfrac{1}{2\pi}
    \int_{\mR}
    \rprod_{k=1}^r
    \Phi_k(\lambda)
    \rd \lambda,
\end{equation}
where $\Phi_k(\lambda):= \int_\mR \re^{-i\lambda t }f_k(t)\rd t$ is the Fourier transform of the kernel function $f_k$. In turn, (\ref{lim}) extends to complex-valued functions $h$ of $r$ complex variables evaluated at the integral operators $\cF_T^{(k)}$:
\begin{align}
\nonumber
    \lim_{T\to +\infty}
    \big(&
        \tfrac{1}{T}
        \Tr h(\cF_T^{(1)}, \ldots, \cF_T^{(r)})
    \big)\\
\label{lim1}
    & =
    \tfrac{1}{2\pi}
    \int_{\mR}
    \Tr
    h(
    \Phi_1(\lambda),
    \ldots,
    \Phi_r(\lambda)
    )
    \rd \lambda,
\end{align}
provided both sides of (\ref{lim1}) use the same extension of $h$ to noncommutative variables (such extensions are, in general, not unique), and the constituent kernel functions satisfy suitable regularity conditions.

The following theorem is concerned with the asymptotic behaviour of the quantity (\ref{lnXi1}), as $T\to +\infty$, and employs the Fourier transforms
\begin{align}
\label{Phi0}
    \Phi(\lambda)
    & :=
    \int_\mR \re^{-i\lambda t }
    P(t)
    \rd t
    =
    F(i\lambda) F(i\lambda)^*,\\
\label{Psi0}
    \Psi(\lambda)
    & :=
    \int_\mR \re^{-i\lambda t }
    \Lambda(t)
    \rd t
    =
    F(i\lambda) J F(i\lambda)^*,
    \qquad
    \lambda \in \mR,
\end{align}
of the covariance and commutator kernels (\ref{P}), (\ref{Lambda}), see also \cite[Eq.~(5.8)]{VPJ_2019a}. Here, $(\cdot)^*:= {{\overline{(\cdot)}}}^\rT$ is the complex conjugate transpose, and
\begin{equation}
\label{F0}
    F(v)
    :=
    S
    (vI_n - A)^{-1}B,
    \qquad
    v \in \mC,
\end{equation}
is the transfer function from the incremented input quantum Wiener process $W$ of the OQHO (\ref{dX}) to the process $Z$ in (\ref{ZX}). Note that $\Phi(\lambda)$ is a complex positive semi-definite Hermitian  matrix, while  $\Psi(\lambda)$  is skew Hermitian for any $\lambda \in \mR$, with $\Phi+i\Psi$ being the Fourier transform  of the quantum covariance kernel $P+i\Lambda$ from (\ref{EZZ}).

\begin{thm}
\label{th:limXi}
Suppose the OQHO (\ref{dX}) is driven by vacuum input fields, the matrix $A$ in (\ref{AB}) is Hurwitz, and the matrix $B$ satisfies (\ref{BJB}).  Also, let the risk sensitivity parameter $\theta>0$ in (\ref{Xi}) satisfy
\begin{equation}
\label{spec1}
    \theta
    \sup_{\lambda \in \mR}
    \lambda_{\max}
    (
        \Phi(\lambda)
        \tanc
        (\theta \Psi(\lambda))
    )
    < 1,
\end{equation}
where the functions $\Phi$, $\Psi$ are given by (\ref{Phi0})--(\ref{F0}). Then the QEF $\Xi_{\theta,T}$, defined by  (\ref{Xi}), (\ref{Q}), has the following infinite-horizon growth rate:
\begin{equation}
\label{Ups}
    \Ups(\theta)
    :=
\lim_{T\to +\infty}
    \big(
        \tfrac{1}{T}
        \ln \Xi_{\theta,T}
    \big)
     =
    -
    \tfrac{1}{4\pi}
    \int_{\mR}
    \ln\det
    D_\theta(\lambda)
    \rd \lambda,
\end{equation}
where
\begin{equation}
\label{D}
    D_\theta(\lambda)
    :=
    \cos(
        \theta \Psi(\lambda)
    ) -
        \theta
        \Phi(\lambda)
        \sinc
        (\theta \Psi(\lambda)),
\end{equation}
and $\sinc z := \frac{\sin z}{z}$ (which is extended  as $\sinc 0 := 1$ by continuity).\hfill$\square$
\end{thm}
\begin{pf}
In the case of one integral operator, the noncommutativity  issue does not arise, and (\ref{lim1}) is directly applicable to the second part of (\ref{lnXi1}) as
\begin{align}
\nonumber
    \lim_{T\to +\infty}
    \big(
        \tfrac{1}{T}
        \Tr \psi(\theta\cL_T )
    \big)
    & =
    \tfrac{1}{2\pi}
    \int_{\mR}
    \Tr \ln\cos(
        \theta \Psi(\lambda)
    )
    \rd \lambda\\
\label{psilim}
    & =
    \tfrac{1}{2\pi}
    \int_{\mR}
    \ln\det \cos(
        \theta \Psi(\lambda)
    )
    \rd \lambda,
\end{align}
where the function $\psi$ is given by (\ref{phipsi}), and use is made of the identity $\Tr \ln N = \ln\det N$ for square matrices $N$, along with the  Fourier transform (\ref{Psi0})
of the commutator kernel (\ref{Lambda}).
Now, in application of (\ref{lim1}) to the first part of (\ref{lnXi1}),  the function $\varphi$ from (\ref{phipsi}) is evaluated at the operator $\theta \cP_T  \cK_{\theta,T}  $ which involves  two noncommuting integral operators $\cP_T $, $\cL_T $ in (\ref{cP}), (\ref{cL}) and the related operator $\cK_{\theta,T}  $ in (\ref{cK}) as
\begin{align}
\nonumber
    \varphi(\theta \cP_T  \cK_{\theta,T}  )
    & =
    -
    \sum_{r=1}^{+\infty}
    \tfrac{1}{r}
    \theta^r
    (\cP_T  \cK_{\theta,T}   )^r\\
\label{phiPK}
    & =
    -
    \sum_{r=1}^{+\infty}
    \tfrac{1}{r}
    \theta^r
    \sum_{k_1, \ldots, k_r = 0}^{+\infty}
    \rprod_{j=1}^r
    \big(
    c_{k_j}
    \theta^{2k_j}
    \cP_T
    \cL_T ^{2k_j}
    \big)
\end{align}
under the condition (\ref{spec}).
Here, use is made of the Maclaurin series expansion $\tanc z = \sum_{k=0}^{+\infty} c_k z^{2k}$ in view of the symmetry of the {\tt tanc} function, with a particular form of the coefficients $c_k \in \mR$ being irrelevant.  Application of (\ref{lim1}) to (\ref{phiPK}) (we omit here the justification of interchangeability of the summation and taking the limit) yields
\begin{align}
\nonumber
     \lim_{T\to +\infty}  &
    \big(
        \tfrac{1}{T}
        \Tr \varphi(\theta\cP_T \cK_{\theta,T}  )
    \big)\\
\nonumber
    & =
    -
    \tfrac{1}{2\pi}
    \sum_{r=1}^{+\infty}
    \tfrac{1}{r}
    \theta^r\!\!\!\!\!\!\!
    \sum_{k_1, \ldots, k_r = 0}^{+\infty}
    \int_\mR\!
    \Tr
    \rprod_{j=1}^r
    \big(
    c_{k_j}
    \theta^{2k_j}
    \Phi(\lambda)
    \Psi(\lambda)^{2k_j}
    \big)
    \rd \lambda\\
\nonumber
    & =
    \tfrac{1}{2\pi}
    \int_{\mR}
    \Tr
    \ln(
    I_n -
        \theta
        \Phi(\lambda)
        \tanc
        (\theta \Psi(\lambda))
    )
    \rd \lambda\\
\label{philim}
    & =
    \tfrac{1}{2\pi}
    \int_{\mR}
    \ln\det(
    I_n -
        \theta
        \Phi(\lambda)
        \tanc
        (\theta \Psi(\lambda))
    )
    \rd \lambda,
\end{align}
where
the Fourier transform (\ref{Phi0}) of the covariance kernel (\ref{P}) is used together with (\ref{Psi0}).  The limit relation (\ref{philim}) holds under the condition (\ref{spec1}) which is a frequency-domain representation of (\ref{spec}).
By combining (\ref{psilim}), (\ref{philim}), it follows that the quantity (\ref{lnXi1}) has the asymptotic growth rate
\begin{align}
\nonumber
\lim_{T\to +\infty}&
    \big(
        \tfrac{1}{T}
        \ln \Xi_{\theta,T}
    \big)
    =
    -
    \tfrac{1}{4\pi}
    \int_{\mR}
    \ln\det(
    I_n -
        \theta
        \Phi(\lambda)
        \tanc
        (\theta \Psi(\lambda))
    )
    \rd \lambda\\
\nonumber
    & -
    \tfrac{1}{4\pi}
    \int_{\mR}
    \ln\det \cos(
        \theta \Psi(\lambda)
    )
    \rd \lambda\\
\label{Xilim}
    = &
    -
    \tfrac{1}{4\pi}
    \int_{\mR}
    \ln\det(
    \cos(
        \theta \Psi(\lambda)
    ) -
        \theta
        \Phi(\lambda)
        \sinc
        (\theta \Psi(\lambda))
    )
    \rd \lambda,
\end{align}
where the identity
$\tanc z \cos z = \sinc z$ is applied to the matrix $\theta \Psi(\lambda)$. In view of (\ref{D}), the relation (\ref{Xilim}) establishes (\ref{Ups}). \hfill$\blacksquare$
\end{pf}

Under the condition (\ref{spec1}), $-\ln\det D_\theta(\lambda)$ is a symmetric function of the frequency $\lambda$ with nonnegative values.
From (\ref{Psi0}), (\ref{F0}), it follows that the Hurwitz property of the matrix $A$,   the nonsingularity  of the matrix $S$ in (\ref{ZX}) and the condition (\ref{BJB}) imply that
\begin{equation}
\label{detPsi}
    \det \Psi(\lambda)
    =
    \tfrac{\det \Pi \det (BJB^\rT) }{|\det (i\lambda - A)|^2}
     \ne 0,
    \qquad
    \lambda \in \mR,
\end{equation}
which makes the extension $\sinc 0 =1$ irrelevant for the evaluation of $        \sinc
        (\theta \Psi(\lambda))
$. However, this extension (and also $\tanc 0 = 1$) plays its role in the limiting classical case, when (\ref{dX}) is an SDE driven by a standard Wiener process $W$ (formally with $J=0$ in (\ref{WWcomm})), and $Z$ in (\ref{ZX}) is a stationary Gaussian diffusion process \cite{GS_2004} with zero mean and the spectral density $\Phi$ in (\ref{Phi0}). In this case,  the function $\Psi$ vanishes, the condition (\ref{spec1}) takes the form
\begin{equation}
\label{class}
\theta <     \theta_0:= \tfrac{1}{\sup_{\lambda \in \mR}
    \lambda_{\max}
    (
        \Phi(\lambda)
    )} = \tfrac{1}{\|F\|_\infty^2}
\end{equation}
in terms of the $\cH_\infty$-norm of the transfer function (\ref{F0}), and the right-hand side of (\ref{Ups}) reduces to the $\cH_\infty$-entropy integral \cite{AK_1981,MG_1990}
\begin{equation}
\label{V}
    V(\theta)
    :=
    -
    \tfrac{1}{4\pi}
    \int_{\mR}
    \ln\det(
    I_n
 -
        \theta
        \Phi(\lambda)
    )
    \rd \lambda.
\end{equation}
In contrast to its classical counterpart, the QEF growth rate (\ref{Ups}) in the quantum case depends  on both functions $\Phi$, $\Psi$ which constitute the ``quantum spectral density'' $\Phi + i\Psi$ of the process $Z$ in (\ref{ZX}). Furthermore, the condition (\ref{spec1}) is substantially nonlinear with respect to  $\theta$ and, unlike (\ref{class}),  does not admit a closed-form representation. However, since {\tt tanc} on the imaginary axis (that is, {\tt tanhc} on the real axis) takes values in the interval $(0,1]$, then
$
    \lambda_{\max}
    (
        \Phi(\lambda)
        \tanc
        (\theta \Psi(\lambda))
    )
    =
    \lambda_{\max}
    (
        \sqrt{\tanc
        (\theta \Psi(\lambda))}
        \Phi(\lambda)
        \sqrt{\tanc
        (\theta \Psi(\lambda))}
    )
    \< \lambda_{\max}(\Phi(\lambda))
$ for any $\lambda \in \mR$, whereby the fulfillment of the classical constraint (\ref{class}) implies (\ref{spec1}).

As a function of $\theta$, the QEF growth rate (\ref{Ups}) plays an important role in quantifying the large deviations of quantum trajectories \cite{VPJ_2018a} and for robustness of the OQHO with respect to state uncertainties described in terms of quantum relative entropy (see \cite[Section~IV]{VPJ_2018b}
 and references therein). More precisely,
\begin{equation}
\label{supP}
    \limsup_{T\to +\infty}
    \big(
        \tfrac{1}{T}
        \ln
        \bP_T([2\alpha T, +\infty))
    \big)
    \<
    \inf_{\theta\>0}
    (
        \Ups(\theta)
        -
        \alpha\theta
    )
\end{equation}
for any $\alpha>0$,
where $\bP_T$ is the probability distribution  of the self-adjoint quantum variable $Q_T$ in (\ref{Q}) \cite{H_2001}. Therefore, (\ref{supP}) provides asymptotic upper bounds for the tail probability distribution of $Q_T$ in terms of the QEF growth rate (\ref{Ups}). Furthermore,
\begin{equation}
\label{supEQ}
  \limsup_{T\to+\infty}
  \Big(
    \tfrac{1}{T}
    \sup_{\sigma \in \fS_{\eps, T}}
    \bE_{\sigma} Q_T
  \Big)
  \<
  2
  \inf_{\theta >0}
  \big(
    \tfrac{1}{\theta}
    (\eps + \Ups(\theta))
  \big),
\end{equation}
where $\bE_\sigma Q_T := \Tr (\sigma Q_T)$ is the expectation of the $\fH_T$-adapted quantum variable $Q_T$ in (\ref{Q}) over a density operator $\sigma$ on the system-field subspace $\fH_T$. Here,  the supremum is taken over the set
\begin{equation}
\label{fS}
    \fS_{\eps, T}
    :=
  \big\{
    \sigma:\
    \bD(\sigma \| \rho_T)\< \eps T
  \big\},
\end{equation}
where the parameter $\eps\> 0$ limits the growth rate of the quantum relative entropy \cite{OW_2010}
\begin{equation}
\label{bD}
  \bD(\sigma \| \rho_T)
  :=
  - \bH(\sigma)
  -\bE_\sigma\ln \rho_T
\end{equation}
of $\sigma$ with respect to $\rho_T:= \fP_T \rho \fP_T$, with $\fP_T$ the orthogonal projection onto $\fH_T$, and
$\bH(\sigma):= -\bE_\sigma \ln \sigma$ is
the von Neumann entropy of $\sigma$; cf. \cite[Eq. (7)]{YB_2009}. The density operator $\sigma$ is interpreted as the actual quantum state, about which it is only known that it belongs to the class (\ref{fS}) of states being not ``too far''  from the reference state $\rho_T$ as a nominal model. In the framework of this quantum statistical uncertainty description, specified by $\eps$ in terms of (\ref{bD}),  the left-hand side of (\ref{supEQ}) is the worst-case quadratic cost growth rate, similar to the robust performance criteria of minimax LQG control \cite{DJP_2000,P_2006,PJD_2000}.

Therefore, for a suitably chosen $\theta>0$,  the minimization of $\Ups(\theta)$ over an admissible range of parameters of the OQHO in the context of risk-sensitive control and filtering problems enhances the large deviations and robust performance bounds (\ref{supP}), (\ref{supEQ}). The computation of these  bounds and the QEF minimization demand techniques for evaluating the functional (\ref{Ups}).

\section{Evaluation of the QEF growth rate}
\label{sec:comp}

One of techniques for computing the QEF growth rate (\ref{Ups})
resembles the homotopy methods for numerical solution of parameter dependent algebraic equations \cite{MB_1985} and exploits the specific dependence of $\Ups(\theta)$ on the risk sensitivity parameter $\theta$. To this end, with the function $D_\theta$ in (\ref{D}),  we associate a function $U_\theta: \mR \to \mC^{n\x n}$ by
\begin{equation}
\label{U}
  U_\theta(\lambda):= -D_\theta(\lambda)^{-1}\d_\theta D_\theta(\lambda)
\end{equation}
for all $\theta >0$ satisfying (\ref{spec1}) (which ensures that $\det D_\theta (\lambda)\ne 0$ for all $\lambda \in \mR$).

\begin{thm}
\label{th:diff}
Under the conditions of Theorem~\ref{th:limXi}, the QEF growth rate $\Ups(\theta)$ in  (\ref{Ups}) satisfies the ODE (more precisely, an integro-differential equation)
\begin{equation}
\label{Ups'}
  \Ups'(\theta)
  =
    \tfrac{1}{4\pi}
    \int_{\mR}
    \Tr U_\theta(\lambda)
    \rd \lambda,
\end{equation}
with the initial condition $\Ups(0)=0$. Here, the function (\ref{U}) is computed as
\begin{equation}
\label{U1}
U_\theta
=
\Psi
    (    \Psi\cos(
        \theta \Psi
        )\!\! -\!\!
        \Phi
        \sin
        (\theta \Psi)
        )^{-1}\!
        (\Phi \cos(\theta \Psi)
    \!\!+\!\!\Psi\sin(\theta \Psi)
    ),
\end{equation}
takes values in the subspace of Hermitian matrices and satisfies a Riccati equation
\begin{equation}
\label{U'}
  \d_\theta U_\theta = \Psi^2 + U_\theta^2
\end{equation}
at any frequency $\lambda\in \mR$,
with the initial condition $U_0 = \Phi$ given by (\ref{Phi0}).\hfill$\square$
\end{thm}
\begin{pf}
The relation (\ref{Ups'}) is obtained from (\ref{Ups}), (\ref{U}) by applying the identity $(\ln\det N)' = \Tr (N^{-1}N')$, where $(\cdot)':= \d_\theta(\cdot)$, so that $(\ln\det D_\theta)' = -\Tr U_\theta$.   Now,
in view of (\ref{detPsi}),  the function $D_\theta$ in (\ref{D})  can be represented as
\begin{equation}
\label{D1}
    D_\theta
    =
    \cos(
        \theta \Psi
    ) -
        \Phi
        \Psi^{-1}
        \sin
        (\theta \Psi)
\end{equation}
for any $\lambda\in \mR$, and hence, its  differentiation with respect to $\theta$ yields
\begin{equation}
\label{D'}
    D_\theta'
    =
    -\Psi
    \sin(
        \theta \Psi
    )
    -
    \Phi
        \cos
        (\theta \Psi).
\end{equation}
Substitution of (\ref{D1}), (\ref{D'}) into (\ref{U}) leads to (\ref{U1}). By differentiating  (\ref{D'}) in $\theta$, it follows that (\ref{D1})
satisfies the linear second-order  ODE
\begin{equation}
\label{D''}
  D_\theta''
  =
    -\Psi^2
    \cos(
        \theta \Psi
    )
    +
    \Phi
    \Psi
        \sin
        (\theta \Psi)
    =
  - D_\theta \Psi^2,
\end{equation}
with the initial conditions $D_0 = I_n$, $D_0' = -\Phi$. Therefore, the differentiation of (\ref{U}) leads to
\begin{equation}
\label{U'1}
    U_\theta'
    =
    -D_\theta^{-1}D_\theta''
    +D_\theta^{-1}D_\theta' D_\theta^{-1}D_\theta'
    =
    \Psi^2 + U_\theta^2,
\end{equation}
where use is made of the relation $(N^{-1})' = -N^{-1}N'N^{-1}$ along with (\ref{D''}). The solution of (\ref{U'1}) inherits the Hermitian property from its initial condition $U_0 = \Phi$, since  $\Psi(\lambda) = -\Psi(\lambda)^*$ in (\ref{Psi0}) for any $\lambda\in \mR$, and $(N^2)^* = N^2$ for Hermitian and skew  Hermitian matrices $N$.
\hfill$\blacksquare$
\end{pf}

The relation (\ref{U}), which links the quadratically nonlinear ODE (\ref{U'}) with the linear ODE (\ref{D''}), can be regarded as a matrix-valued analogue of the Hopf-Cole transformation \cite{C_1951,H_1950} converting the viscous Burgers equation to the heat (or diffusion) equation.  We also mention an analogy between (\ref{U}) and the logarithmic transformation in the context of dynamic programming
equations for stochastic control  \cite{F_1982} (see also \cite{VP_2010}).

The right-hand side of (\ref{Ups'}) can be evaluated by numerical integration over the frequency axis and used for computing (\ref{Ups}) as
\begin{equation}
\label{homo}
    \Ups(\theta)
    =
    \int_0^\theta \Ups'(v)\rd v
    =
    \tfrac{1}{4\pi}
    \int_{\mR \x [0,\theta]}
    \Tr U_v(\lambda)
    \rd \lambda \rd v.
\end{equation}
In particular, (\ref{Ups'}) yields
\begin{equation}
\label{Ups'0}
    \Ups'(0)
    =
    \tfrac{1}{4\pi}
    \int_\mR
    \Tr \Phi(\lambda)
    \rd \lambda
    =
    \tfrac{1}{2}
    \|F\|_2^2 =
    \tfrac{1}{2}
    \bE(Z(0)^\rT Z(0)),
\end{equation}
which, in accordance with (\ref{Xi}),  reproduces the LQG cost for the process $Z$ in (\ref{ZX}) for the stable OQHO in the invariant Gaussian state. In (\ref{Ups'0}), we have also used the $\cH_2$-norm of the transfer function (\ref{F0}) which factorizes the spectral density (\ref{Phi0}). In addition to its role for the computation of $\Ups$, the function $\Ups'$ admits the following representation (see also \cite[Theorem 1]{VPJ_2018a}):
\begin{equation}
\label{Ups'1}
  \Ups'(\theta)
  =
  \tfrac{1}{2}
  \lim_{T\to +\infty}
  \big(
    \tfrac{1}{T}
    \bE_{\theta, T} Q_T
  \big),
\end{equation}
where $\bE_{\theta, T} \zeta := \Tr(\rho_{\theta, T}\zeta)$ is the quantum expectation over a modified density operator $\rho_{\theta, T}:= \frac{1}{\Xi_{\theta, T}}\re^{\frac{\theta}{4}Q_T} \rho \re^{\frac{\theta}{4}Q_T}$.  Therefore, (\ref{Ups'1}) relates $\Ups'$ to the asymptotic growth rate of the weighted average of the quantum variable $Q_T$ in (\ref{Q}) rather than its exponential moment.

\begin{figure}[htbp]
\begin{center}
\includegraphics[width=5cm]{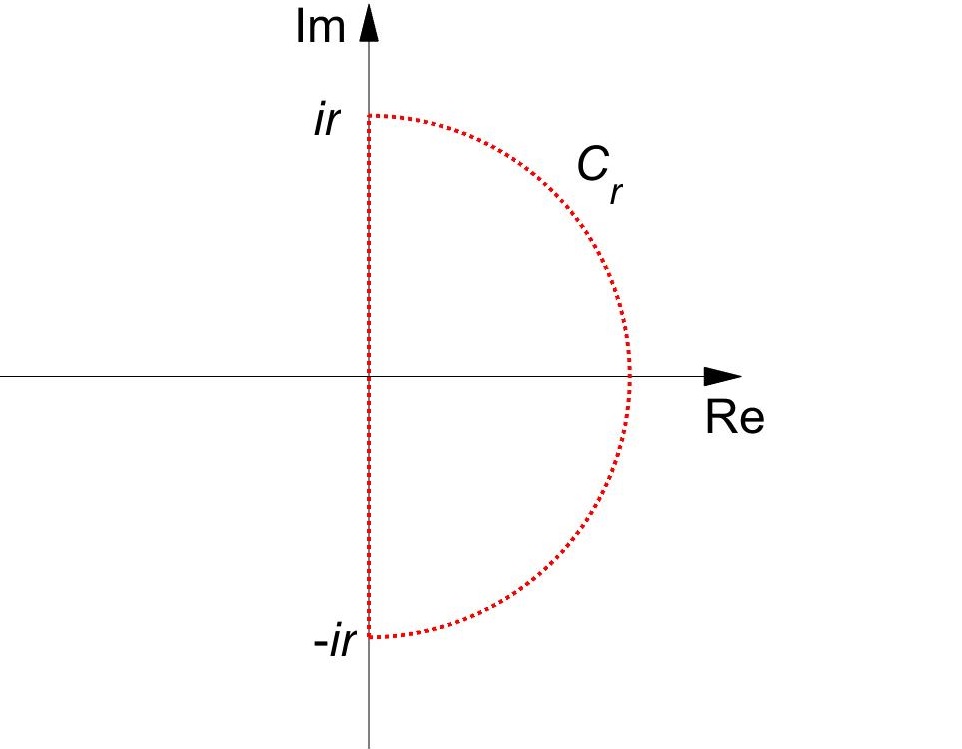}
\caption{The counterclockwise oriented contour $C_r$ in (\ref{Ups1}) consisting of an arc of radius $r$ (centered at the origin) and a line segment of the imaginary axis with the endpoints $\pm ir$.}
\label{fig:cont}
\end{center}
\end{figure}

Another approach to evaluating the QEF growth rate (\ref{Ups}) is provided by contour integration. More precisely, consider  the $\mC^{n\x n}$-valued function
\begin{equation}
\label{E}
    E_\theta(s)
    :=
    \cos(
        \theta \mho(s)
    ) -
        \theta
        \Gamma(s)
        \sinc
        (\theta \mho(s)),
\end{equation}
which is defined in terms of the rational (and hence, meromorphic) functions
\begin{align}
\label{Gamma}
    \Gamma(s)
     & :=
    F(s) F(-s)^\rT,\\
\label{mho}
    \mho(s)
    & :=
    F(s) J F(-s)^\rT,
    \qquad
    s \in \mC,
\end{align}
associated with the transfer function (\ref{F0}). Since (\ref{E})--(\ref{mho}) are related to (\ref{D}), (\ref{Phi0}),  (\ref{Psi0}) as $D_\theta(\lambda) = E_\theta(i\lambda)$, $\Phi(\lambda) = \Gamma(i\lambda)$, $\Psi(\lambda) = \mho(i\lambda)$ for all $\lambda \in \mR$, then (\ref{Ups}) admits the representation
\begin{align}
\nonumber
    \Ups(\theta)
    & =
      -
    \tfrac{1}{4\pi i}
    \int_{i\mR}
    \ln\det E_{\theta}(s)
    \rd s\\
\label{Ups1}
    & =
    \tfrac{1}{4\pi i}
    \lim_{r\to +\infty}
    \oint_{C_r}
    \ln\det E_{\theta}(s)
    \rd s,
\end{align}
where the last integral is over the counterclockwise oriented contour in Fig.~\ref{fig:cont}.
Here, use is made of the asymptotic behaviour
\begin{equation}
\label{Easy}
    E_\theta(s) = I_n
    +
    \tfrac{\theta}{s^2}
    SBB^\rT S
    + o(s^{-2})
\end{equation}
of the function (\ref{E}), as $s\to \infty$,
due to the transfer function $F$ in (\ref{F0}) being strictly proper. Therefore,
the contribution from the semicircular part $C_r\bigcap \mC_+ = \{s \in \mC_+:\ |s|=r\} $
of the contour $C_r$ in (\ref{Ups1}) indeed vanishes asymptotically:
$    \int_{C_r\bigcap \mC_+}
    \ln\det E_{\theta}(s)\rd s \sim 2i\theta \Tr(\Pi BB^\rT)\frac{1}{r}$,
as $r\to +\infty$,
where $\mC_+ := \{s \in \mC:\ \Re s >0\}$ is the open right half-plane,    and the relation $\det (I_n + N) = 1 + \Tr N + o(N)$, as $N \to 0$, is used together with the identity $\Tr(S BB^\rT S) =   \Tr(S^2 BB^\rT) = \Tr(\Pi BB^\rT)$ which follows from the structure of the matrix $S$ in (\ref{ZX}). However, application of the residue theorem \cite{S_1992} to (\ref{Ups1})
is complicated by the nature of singularities of the function $\det E_\theta$ (considered in $\mC_+$),
which will be demonstrated in Section~\ref{sec:one}. Note that  the corresponding function $\det (I_n - \theta \Phi)$ in the classical counterpart (\ref{V}) is rational, thus simplifying the evaluation of the integral. This observation can be combined with the Maclaurin series expansions of the trigonometric functions, which allows (\ref{D}) to be approximated as
\begin{align}
\nonumber
    D_\theta
    & =
    I_n - \tfrac{1}{2}\theta^2 \Psi^2 - \theta \Phi (I_n - \tfrac{1}{6}\theta^2 \Psi^2)
    +
    o(\theta^3)\\
\label{Dprox}
    & =
    I_n - \theta \Phi
    -
    \tfrac{1}{2}\theta^2
    (I_n - \tfrac{\theta}{3}\Phi)\Psi^2
    +
    o(\theta^3)
\end{align}
as $\theta \to 0$. Substitution of (\ref{Dprox}) into (\ref{Ups}) leads to the approximate computation of the QEF growth rate as a perturbation of its classical counterpart (\ref{V}):
\begin{align}
\nonumber
  \Ups(\theta)
  = &
  V(\theta)\\
\nonumber
  & +
  \tfrac{\theta^2}{8\pi}
  \int_\mR
  \Tr
  ((I_n - \theta\Phi(\lambda))^{-1}(I_n - \tfrac{\theta}{3}\Phi(\lambda))\Psi(\lambda)^2)
  \rd \lambda\\
\label{VUps}
    & + o(\theta^3),
    \qquad
    {\rm as}\
    \theta \to 0.
\end{align}
Since the integrand in (\ref{VUps}) is a rational function of the frequency $\lambda$, whose continuation to the closed right half-plane $(i\mR) \bigcup \mC_+$ has no poles on the imaginary axis under the condition (\ref{class}), the correction term is amenable to calculation via its residues in $\mC_+$. In view of $\Psi(\lambda)^2\prec 0$ for all $\lambda \in \mR$, the relation (\ref{VUps}) also implies that $\Ups(\theta)< V(\theta)$ for all sufficiently small $\theta >0$.

\section{QEF growth rate for a one-mode OQHO}
\label{sec:one}

As an example,
consider
a one-mode ($n=2$) OQHO with the conjugate position-momentum pair (\ref{zeta}) as the system variables, so that
    $X =
    {\scriptsize\begin{bmatrix}
        q\\
        p
    \end{bmatrix}}$,
and the CCR matrix (\ref{XCCR}) reduces to $\Theta = \frac{1}{2}\bJ$, with $\bJ$ given by (\ref{bJ}). In this case,
\begin{equation}
\label{MJM}
  M^\rT J M = \mu \bJ
\end{equation}
for some $\mu \in \mR$, regardless of a particular structure of the coupling matrix $M \in \mR^{m\x 2}$.  In what follows, it is assumed that $\mu>0$ and the energy matrix $R$ is positive definite.  Then the matrix $A \in \mR^{2\x 2}$ in (\ref{AB}) can be computed as \cite[Theorem~1]{VJP_2019}
\begin{equation}
\label{AR1}
    A
    =
    R^{-1/2}(\nu \bJ  - \mu I_2)\sqrt{R},
    \qquad
    \nu
    := \sqrt{\det R}
\end{equation}
and is Hurwitz (its eigenvalues are $-\mu \pm i \nu$). The corresponding matrix $B = \bJ M^\rT$ in (\ref{AB}) satisfies
\begin{equation}
\label{BJBmu}
    BJB^\rT = -\bJ M^\rT J M \bJ = -\mu \bJ^3 = \mu \bJ
\end{equation}
in view of (\ref{MJM}) and the property $\bJ^2 = -I_2$.
Due to the similarity transformation in (\ref{AR1}), the transfer function $F$ in (\ref{F0}) takes the form
\begin{align}
\nonumber
    F(s)
    & =
    S
    (sI_2 - R^{-1/2}(\nu \bJ  - \mu I_2)\sqrt{R})^{-1}B\\
\label{F1}
    & =
    S R^{-1/2}
    ((s+\mu)I_2 - \nu \bJ)^{-1}\sqrt{R}B.
\end{align}
By substituting (\ref{F1}) into (\ref{mho}) and taking (\ref{BJBmu}) into account, it follows that
\begin{align}
\nonumber
    \mho(s)
    = &
    \mu
    S R^{-1/2}
    ((s+\mu)I_2 - \nu \bJ)^{-1}\sqrt{R}\bJ \sqrt{R}    \\
\nonumber
    & \x
    ((\mu-s)I_2 + \nu \bJ)^{-1}
    R^{-1/2} S\\
\nonumber
    = &
    \mu \nu
    S R^{-1/2}
    ((s+\mu)I_2 - \nu \bJ)^{-1}\bJ     \\
\nonumber
    & \x
    ((\mu-s)I_2 + \nu \bJ)^{-1}
    R^{-1/2} S    \\
\nonumber
    = &
    \tfrac{\mu \nu}{((s+\mu)^2 + \nu^2)((\mu-s)^2 + \nu^2)}\\
\nonumber
    & \x
    S R^{-1/2}
    ((s+\mu)I_2 + \nu \bJ)\bJ     \\
\nonumber
    & \x
    ((\mu-s)I_2 - \nu \bJ)
    R^{-1/2} S\\
\nonumber
    = &
    \tfrac{\mu \nu}{((s+\mu)^2 + \nu^2)((\mu-s)^2 + \nu^2)}\\
\nonumber
    & \x
    S R^{-1/2}
    ((s+\mu)\bJ - \nu I_2)     \\
\nonumber
    & \x
    ((\mu-s)I_2 - \nu \bJ)
    R^{-1/2} S    \\
\nonumber
    = &
    \tfrac{\mu \nu}{((s+\mu)^2 + \nu^2)((\mu-s)^2 + \nu^2)}\\
\label{mho1}
    & \x
    S R^{-1/2}
    (2\nu s I_2 + (\mu^2+\nu^2-s^2)\bJ)
    R^{-1/2} S,
\end{align}
where use is made of the identity $NJN^\rT = \bJ\det N $,  which holds for any $(2\x 2)$-matrix $N$ and implies that $\sqrt{R} \bJ \sqrt{R} = \bJ \det \sqrt{R} = \nu \bJ$ in view of (\ref{AR1}).
Now, let the weighting matrix $\Pi$ in (\ref{Q}) coincide with the energy matrix:
\begin{equation}
\label{PiR}
  \Pi:= R,
\end{equation}
so that the quantum variable $Q_T$ is related to the time-varying Hamiltonian (\ref{H}) as
\begin{equation}
\label{QH}
    Q_T
    =
    \int_0^T
    X(t)^\rT R X(t)\rd t
    =
    2
    \int_0^T
    H(t)\rd t.
\end{equation}
The resulting QEF
\begin{equation}
\label{XiH}
    \Xi_{\theta, T}
    =
    \bE \re^{\theta \int_0^T
    H(t)\rd t}
\end{equation}
in  (\ref{Xi}) is the moment-generating  function for the ``total'' energy of the OQHO over the time interval $[0,T]$. In this case, $S R^{-1/2} = I_2$ in view of (\ref{PiR}), so that (\ref{F1}), (\ref{mho1}) reduce to
\begin{align}
\label{F2}
    F(s)
    & =
    ((s+\mu)I_2 - \nu \bJ)^{-1}\sqrt{R}B,\\
\label{mho2}
    \mho(s)& =
    a(s) I_2 + b(s) \bJ,
\end{align}
where $a, b: \mC\to \mC$ are rational functions given by
\begin{equation}
\label{ab}
  {\small\begin{bmatrix}
        a\\
        b
  \end{bmatrix}}
  =
  \tfrac{\mu \nu}{((s+\mu)^2 + \nu^2)((\mu-s)^2 + \nu^2)}
  {\small\begin{bmatrix}
        2\nu s \\
        \mu^2+\nu^2-s^2
  \end{bmatrix}},
  \qquad
  s \in \mC.
\end{equation}
Since the right-hand side of (\ref{mho2}) is a linear combination of the  matrices $I_2$, $\bJ$, the trigonometric functions of $\mho(s)$ in (\ref{E}) can be computed as
\begin{align}
\label{cos}
  \cos(\theta \mho)
  & =
  \cos(\theta a) \cosh (\theta b) I_2
  -
  \sin(\theta a) \sinh(\theta b) \bJ,\\
\label{sin}
  \sin(\theta \mho)
  & =
  \sin(\theta a) \cosh (\theta b) I_2
  +
  \cos(\theta a) \sinh(\theta b) \bJ,
\end{align}
where use is made of the identities
$    \cos(z \bJ) = \cosh z I_2$ and
$
    \sin(z \bJ) = \sinh z  \bJ
$ for all $z \in \mC$. Each of the functions $a$, $b$ in (\ref{ab}) has four poles at
\begin{equation}
\label{s1234}
    s_{1,2}:= \mu \pm \nu  i,
    \qquad
    s_{3,4}:= -\mu \pm\nu i.
\end{equation}
The residue of the function $\mho$ in (\ref{mho2}) at each of these poles is a singular matrix:
\begin{equation}
\label{Resmho}
    \det \big(\Res_{s=s_k} \mho(s)\big) = 0,
    \qquad
    k =1,2,3,4,
\end{equation}
and a similar property holds for the function $\Gamma$ in (\ref{Gamma}) associated with (\ref{F2}).  In view of (\ref{cos}), (\ref{sin}), the meromorphic function $\mho$ enters (\ref{E}) through the rational functions $a$, $b$ in composition with the trigonometric functions, which  can manifest an exponential growth or oscillatory behaviour (or both)  depending on the direction in the complex plane. It is for this reason that (\ref{cos}), (\ref{sin}), (\ref{Resmho}) make nontrivial contributions to the singularity of the function $\det E_\theta$ at the points (\ref{s1234}) for computing the QEF growth rate, associated with (\ref{QH}), (\ref{XiH}). Because of the complexity of the resulting expressions, these issues will be discussed in more detail elsewhere.

\section{Numerical example with a two-mode OQHO }
\label{sec:num}

\begin{figure}[htbp]
\begin{center}
\includegraphics[width=8.9cm]{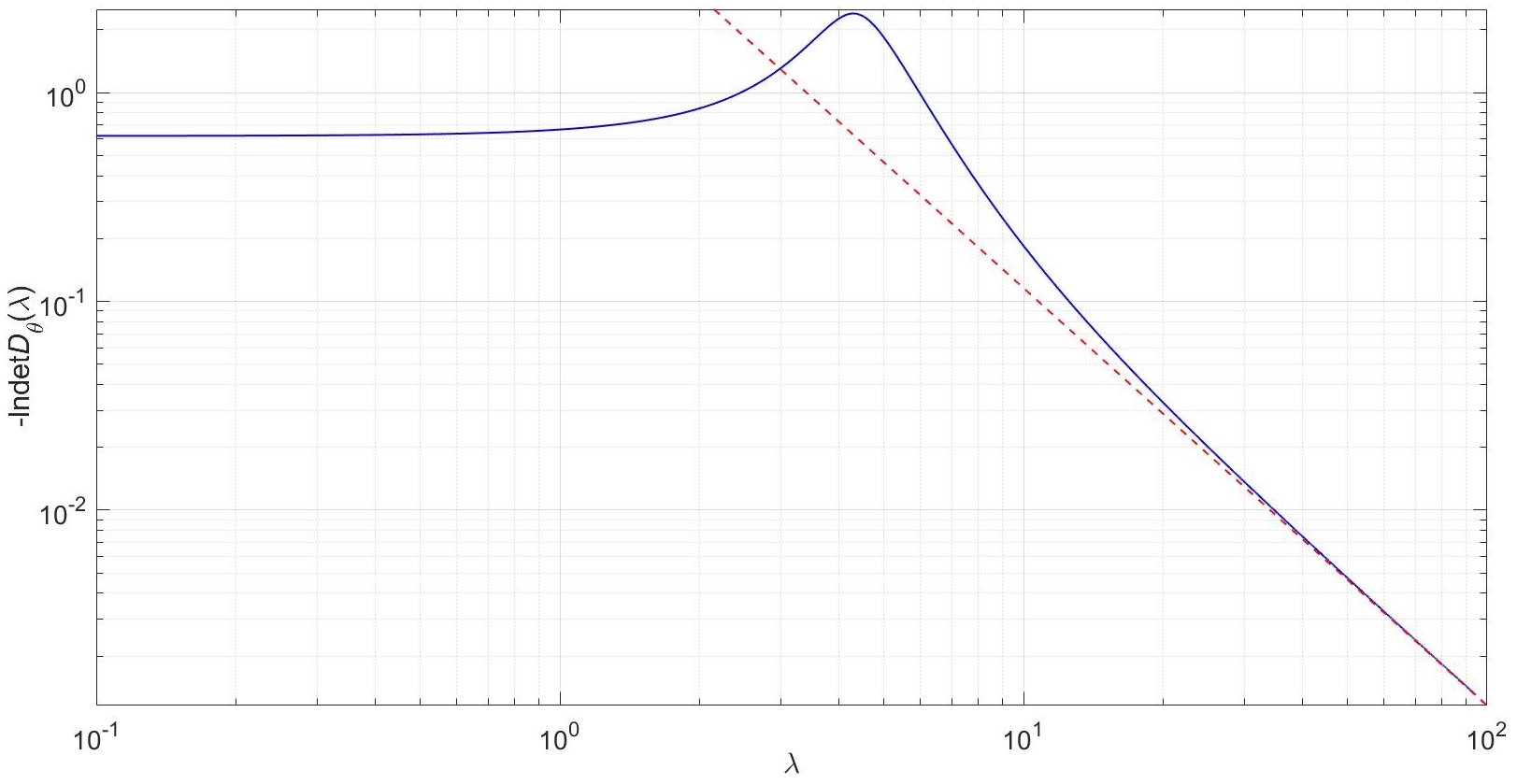}
\caption{The graph of the function $-\ln\det D_\theta(\lambda)$ for positive frequencies $\lambda>0$ (solid line). The dashed line represents the high-frequency asymptote $\frac{\theta}{\lambda^2} \Tr(\Pi BB^\rT)$, as $\lambda\to \infty$, following from (\ref{Easy}). }
\label{fig:lndetD}
\end{center}
\end{figure}

\begin{figure}[htbp]
\begin{center}
\includegraphics[width=8.9cm]{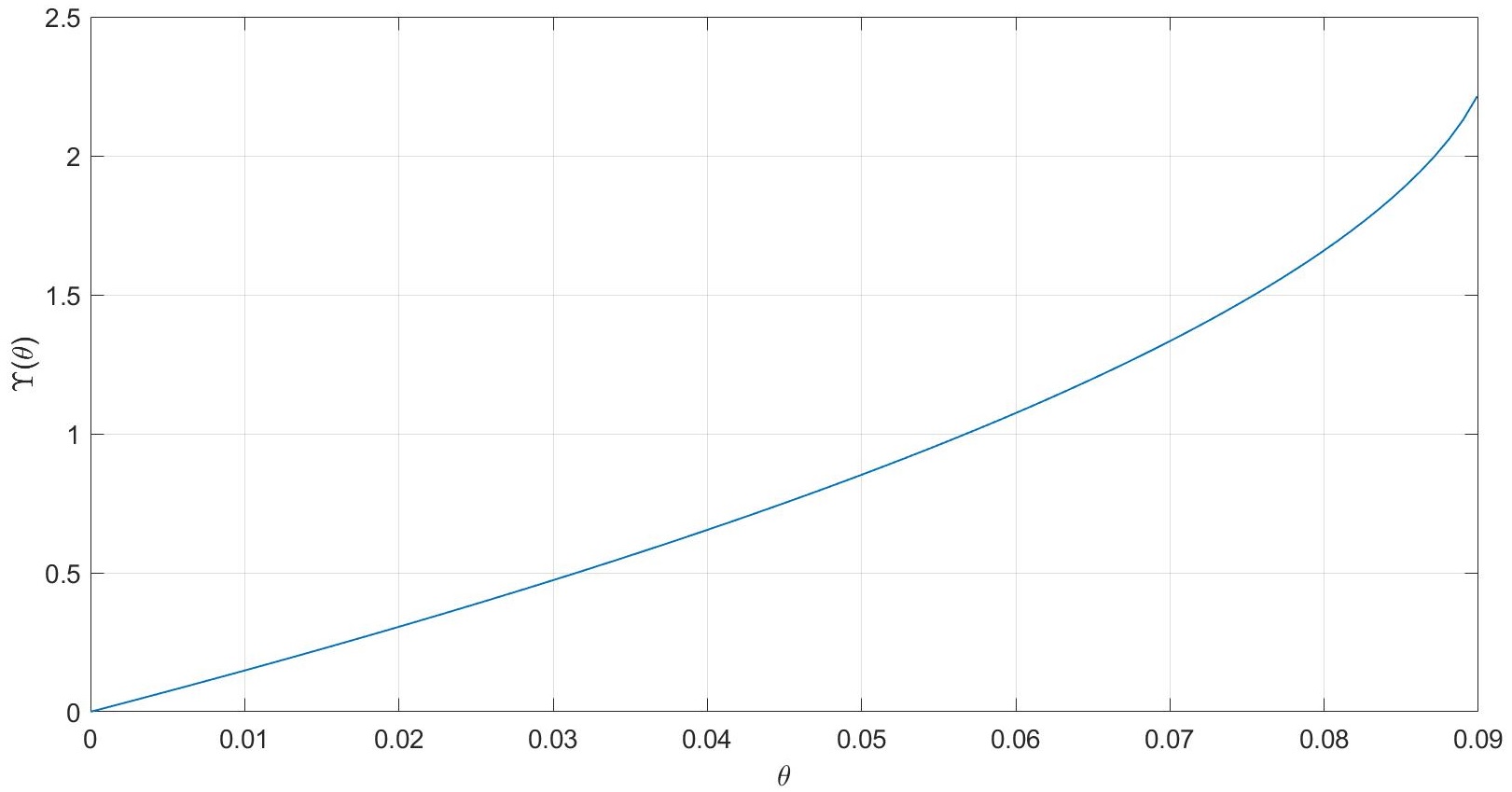}
\caption{The graph of the QEF growth rate (\ref{Ups}) as a function of the risk sensitivity parameter $\theta$. }
\label{fig:Ups}
\end{center}
\end{figure}

Consider a two-mode OQHO, whose $n=4$ system variables consist of two position-momentum pairs, which have the CCR matrix  $\Theta:= \frac{1}{2}\bJ\ox I_2$ and are driven by $m=6$ quantum Wiener processes. The state-space matrices $A$, $B$ in  (\ref{AB}) and the weighting matrix $\Pi$ in (\ref{Q}) are given by
\begin{align}
\label{AA}
A & :=
   {\small\begin{bmatrix}
   -5.8100 &  -1.6357 &   0.2062 &  -3.1331\\
    4.0006 &   0.1377 &   5.3578 &  -0.5514\\
    1.1223 &  -3.0351 &  -5.7830 &   4.4308\\
    2.7957 &  -0.8671 &  -2.2443 &  -0.0737
   \end{bmatrix}},\\
\nonumber
B & :=
   {\small\begin{bmatrix}
   -0.4698  &  0.5026 &   1.9107 &  -1.0020 &   1.8676 &  -1.0523\\
    0.8036  & -0.0727 &  -1.9520 &   2.4997 &  -1.2066 &  -0.7074\\
   -0.1061  & -0.1776 &   0.9175 &  -0.3621 &  -0.2116 &   2.3771\\
   -2.2158  & -1.3753 &  -1.2109 &  -0.8576 &   0.3423 &   1.1991
   \end{bmatrix}},\\
\nonumber
\Pi
    & :=
   {\small\begin{bmatrix}
    3.2123 &   3.5111 &   1.3912 &  -1.8097\\
    3.5111 &  10.6258 &   3.7561 &  -3.7850\\
    1.3912 &   3.7561 &   3.3244 &  -0.5456\\
   -1.8097 &  -3.7850 &  -0.5456 &   1.9349
   \end{bmatrix}}.
\end{align}
In this example, the threshold (\ref{class}) is $\theta_0 = 0.0908$. The graph of the function $-\ln\det D_{\theta}$ from (\ref{Ups}), (\ref{D}) for $\theta = 0.9\theta_0 = 0.0817$  is shown in Fig.~\ref{fig:lndetD} along with its high-frequency asymptote.
The results of numerical computation of the QEF growth rate using (\ref{homo}) and Theorem~\ref{th:diff} are shown in Fig.~\ref{fig:Ups}.
The numerical integration (over positive frequencies in view of the symmetry of the integrand)
employed a combination of a mesh of step size $0.005$  for a low-frequency range $[0,100]$  and the high-frequency asymptote for $\lambda >100$. The choice of the cutoff frequency was based on the spectrum $\{-3.4734 \pm 2.6849i, -2.2911 \pm 4.1584i\}$ and the operator norm $\|A\| = 9.4475$ of the matrix (\ref{AA}). The integration over $\theta$  was carried out with step size $0.01 \theta_0 = 9.08 \x  10^{-4}$.

\section{Conclusion}
\label{sec:conc}

We have established a frequency-domain formula for the infinite-horizon QEF growth rate at the invariant Gaussian state of a stable multimode OQHO driven by multichannel vacuum fields. This representation involves the  quantum spectral density, whose parts are expressed in terms of the  transfer function of the system. We have obtained a  differential equation for the QEF growth rate as a function of the risk sensitivity parameter and outlined its computation using a homotopy technique. A contour integration approach has also been discussed for this purpose along with a more complicated nature of singularities in compositions of trigonometric and matrix-valued rational functions. The latter requires the development of novel spectral factorization techniques (and state-space equations) for this class of computational problems which go beyond the standard application of the residue theorem to rational functions.
The results of the paper provide a
solution of the risk-sensitive robust performance analysis problem  for linear quantum stochastic systems, which will be applied in future publications to coherent and measurement-based control and filtering settings for such systems.

\end{document}